\documentclass[useAMS,usenatbib]{mn2e}
\usepackage{times}
\usepackage{graphicx}
\usepackage{epsfig}
\usepackage[T1]{fontenc}
\usepackage{ifpdf}
\usepackage[varg]{txfonts}

\newcommand{\Msun}      {\mbox{\,$M_{\mathord\odot}$}}

\def\kms{km\,s$^{-1}$}

\title[Identifying IGR J14091--6108 as a magnetic CV]{Identifying 
IGR J14091--6108 as a magnetic CV with a massive white dwarf using 
X-ray and optical observations}

\author[Tomsick et al.]{
John A. Tomsick$^{1}$\thanks{E-mail: jtomsick@ssl.berkeley.edu (JAT)},
Farid Rahoui$^{2,3}$,
Roman Krivonos$^{4}$,
Ma\"ica Clavel$^{1}$,
Jay Strader$^{5}$, \newauthor
and Laura Chomiuk$^{5}$
\\
$^{1}$Space Sciences Laboratory, 7 Gauss Way, University of 
California, Berkeley, CA 94720-7450, USA\\
$^{2}$European Southern Observatory, Karl Schwarzschild-Strasse
2, 85748 Garching bei Munchen, Germany\\
$^{3}$Department of Astronomy, Harvard University, 60 Garden Street, 
Cambridge, MA 02138, USA\\
$^{4}$Space Research Institute, Russian Academy of Sciences, 
Profsoyuznaya 84/32, 117997 Moscow, Russia\\
$^{5}$Department of Physics and Astronomy, Michigan State University, 
East Lansing, MI 48824, USA}

\begin{document}


\def\lsim{\mathrel{\lower .85ex\hbox{\rlap{$\sim$}\raise
.95ex\hbox{$<$} }}}
\def\gsim{\mathrel{\lower .80ex\hbox{\rlap{$\sim$}\raise
.90ex\hbox{$>$} }}}

\pagerange{\pageref{firstpage}--\pageref{lastpage}} \pubyear{2014}

\maketitle

\label{firstpage}

\begin{abstract}

\noindent
IGR~J14091--6108 is a Galactic X-ray source known to have an iron emission
line, a hard X-ray spectrum, and an optical counterpart.  Here, we report
on X-ray observations of the source with {\em XMM-Newton} and {\em NuSTAR} 
as well as optical spectroscopy with ESO/VLT and NOAO/SOAR.  In the X-rays, 
this provides data with much better statistical quality than the previous 
observations, and this is the first report of the optical spectrum.  Timing 
analysis of the {\em XMM} data shows a very significant detection of 
$576.3\pm 0.6$\,s period.  The signal has a pulsed fraction of $30\pm 3$\% 
in the 0.3--12\,keV range and shows a strong drop with energy.  The optical 
spectra show strong emission lines with significant variability in the lines 
and continuum, indicating that they come from an irradiated accretion disk.  
Based on these measurements, we identify the source as a magnetic Cataclysmic 
Variable of Intermediate Polar (IP) type where the white dwarf spin period 
is 576.3\,s.  The X-ray spectrum is consistent with the continuum emission 
mechanism being due to thermal Bremsstrahlung, but partial covering 
absorption and reflection are also required.  In addition, we use the
IP mass (IPM) model, which suggests that the white dwarf in this system 
has a high mass, possibly approaching the Chandrasekhar limit.

\end{abstract}


\section{Introduction}

The hard X-ray imaging by the {\em International Gamma-Ray Astrophysics Laboratory 
(INTEGRAL)} satellite \citep{winkler03} has led to the discovery of a large number 
of new or previously poorly studied ``IGR'' sources.  The most recent published 
catalogs of 17--100\,keV sources detected by {\em INTEGRAL} include more than 
900 sources for the whole sky \citep{bird16} and $\sim$400 sources within $17.5^{\circ}$ 
of the Galactic plane \citep{krivonos12}.  A large fraction of the sources have been 
identified as Active Galactic Nuclei (AGN), Cataclysmic Variables (CVs), High-Mass 
X-ray Binaries (HMXBs), Low-Mass X-ray Binaries (LMXBs), and Pulsar Wind Nebulae 
(PWNe), but 23\% of the 939 sources listed in \cite{bird16} are still unidentified, 
in part due to the difficulty in finding counterparts at other wavelengths with the 
$\sim$1$^{\prime}$--4$^{\prime}$ {\em INTEGRAL} position uncertainties.  In an effort 
to characterize the populations of hard X-ray sources in the Galaxy, we have been 
performing follow-up observations of sources in the Galactic Plane to identify 
the natures of as many IGR sources as possible.

During 2013--2015, we observed ten IGR sources with the {\em Chandra X-ray 
Observatory}, and the results were reported in \cite{tomsick15} and \cite{tomsick16}.  
In some cases, the information from the {\em Chandra} observation, including accurate 
source positions that provided optical or IR identifications, immediately led to a
determination of the nature of the source.  For example, IGR~J04059+5416 and
IGR~J08297--4250 were identified as AGN \citep{tomsick15}, and IGR~J18088--2741
was identified as a CV \citep{tomsick16}.  IGR~J14091--6108 was one of the
sources observed in this program, but the {\em Chandra} observation did not
allow us to definitively determine its nature, and we selected it for further
study as described in this work.

IGR~J14091--6108 was discovered when it was detected in the {\em INTEGRAL} 9-year 
Galactic Hard X-ray Survey \citep{krivonos12}.  A {\em Swift} X-ray counterpart was 
found by \cite{landi12} and then the source was observed with {\em Chandra}, leading 
to the detection of a strong iron K$\alpha$ emission line in the X-ray spectrum and 
the identification of an optical/IR counterpart \citep{tomsick14_atel,tomsick16}.
{\em Chandra} showed that the source has a hard power-law continuum with a photon
index of $\Gamma = 0.6\pm 0.4$, suggesting an accreting compact object with a 
high magnetic field strength \citep{tomsick16}.  While we favored a magnetic CV, 
we did not rule out the possibility that the source is an HMXB.  In this work, we 
use new X-ray and optical observations to study IGR~J14091--6108 in more detail.  

\section{Observations and Data Reduction}

Table~\ref{tab:obs} lists the observations that we used, including simultaneous
X-ray observations with the {\em X-ray Multi-Mirror Mission (XMM-Newton)} and
the {\em Nuclear Spectroscopic Telescope Array (NuSTAR)} as well as optical 
spectroscopy with the Very Large Telescope (VLT\footnote{We carried out the 
observations under ESO program identifier 095.D-0972(A).}) and with the Southern
Astrophysical Research Telescope (SOAR).  The X-ray observations occurred
on 2015 July 20-21, and the observation identifiers (ObsIDs), exact start
and stop times, and exposure times are provided in Table~\ref{tab:obs}.  
The X-ray data provide a large improvement in the statistical quality 
over the previous {\em Chandra} and {\em INTEGRAL} observations to give
better constraints on the spectrum and to allow for a sensitive search
for periodic signals. The optical data provide a first look at the spectrum
of the source.  In the following, we describe how the data from each facility 
were reduced.

\subsection{{\em XMM}}

For the EPIC/pn \citep{struder01} and EPIC/MOS \citep{turner01} instruments, we 
reduced the data using the {\em XMM} Science Analysis Software v14.0.0 to make
images, light curves, and spectra.  For pn, we made a full-field 10--12\,keV light 
curve to look for times of high background, and filtered these times out for the 
final data products.  We followed the same procedure for MOS and obtained slightly 
more time on-source (28.9\,ks for MOS compared to 24.7\,ks for pn).  In addition
to the time filtering, we used the standard event filtering described in procedures 
that are available on-line\footnote{see http://xmm.esac.esa.int/sas/current/documentation/threads/}.
We extracted source counts from a circular region with a radius of $40^{\prime\prime}$ 
centered on IGR~J14091-6108 and estimated the background using a rectangular region
on parts of the detectors with no sources.  After background subtraction, the
0.3--12\,keV count rates for pn, MOS1, and MOS2 are $0.207\pm 0.003$, $0.059\pm 0.015$, 
and $0.065\pm 0.016$\,c/s, respectively.  We included 1\% systematic errors on the 
pn and MOS source spectra when fitting the energy spectra to account for calibration 
uncertainties.  We examined the data from the Reflection Grating Spectrometer (RGS), 
but the count rate for this instrument is too low to be useful.  

\subsection{{\em NuSTAR}}

For {\em NuSTAR} \citep{harrison13}, we reduced the data using HEASOFT v6.17, which
includes NUSTARDAS v1.5.1.  The calibration files are from the 2015 March 16 version
of the calibration database (CALDB).  We ran {\ttfamily nupipeline} to produce 
event lists for the two {\em NuSTAR} instruments: focal plane modules A and B
(FPMA and FPMB).  We examined the full band images, and extracted light curves
and spectra with {\ttfamily nuproducts} using a circular source region with a 
radius of $60^{\prime\prime}$.  Although the {\em NuSTAR} images do not show any
significant stray light, spatial variations in the background are visible, which
is expected.  Thus, we estimated the background using {\ttfamily nuskybgd} 
\citep{wik14}, which is a package of IDL programs that sample the background
over the field of view outside of the source region to model the spatial variations.  
We produced the model and then used it to estimate the background in the source
region.  We included 4\% systematic errors on the background spectrum.  After 
background subtraction, the 3--79\,keV source count rates are $0.076\pm 0.002$ and 
$0.075\pm 0.002$\,c/s for FPMA and FPMB, respectively.  Finally, when producing
the final spectra for fitting, we grouped the energy bins to give a significance
of greater than 5-$\sigma$.

\subsection{Optical Spectroscopy}

We obtained low-resolution spectroscopy of the IGR~J14091--6108 counterpart
identified in \cite{tomsick16}, which is CXOU~J140846.0--610754 and 
VVV~J140845.99--610754.1.  We used the FORS2 instrument with the 300V and 300I 
grisms combined with the GC435 and OG590 filters, respectively. In both cases, the 
slit-width was set to 1\arcsec, giving a $R\sim600$ average resolution. Atmospheric 
conditions were medium-to-good, with a thin sky, seeing at 500~nm in the range 
0\,\farcs6-0\,\farcs8, and an airmass between 1.1 and 1.3. The integration time of 
each individual frame was set to 600\,s and 500\,s for the 300V, and 300I grisms, 
respectively, and two exposures were taken in each grism. The A0V spectro-photometric 
standard star CD-32~9927 was observed in similar conditions for flux-calibration. 
We reduced the data using the dedicated pipeline (v.~5.1.4) implemented in the ESO 
data reduction environment {\tt Reflex}~v.~2.6, following the standard steps for 
optical spectroscopy reduction to produce a cleaned, flatfielded, wavelength- and 
flux-calibrated 1D spectrum.

We also obtained several optical spectra on UT 2015 January 15 using the Goodman 
High-Throughput Spectrograph \citep{clemens04} on the SOAR 4.1-m telescope. 
All data were taken with a $1.03\arcsec$ slit. We obtained one 600 sec exposure 
using the 400 l mm$^{-1}$ grating (wavelength coverage $\sim$3000--7000 \AA; 
resolution 5.7 \AA) and two 600 sec exposures using the 1200 l mm$^{-1}$ grating 
(wavelength coverage 5480--6740 \AA; resolution 1.7 \AA).  These data were 
wavelength calibrated with an FeAr arc lamp and were bias corrected using the
overscan region.  The spectra were optimally extracted using routines in IRAF.

\section{Results}

\subsection{X-ray Timing}

\begin{figure*}
\begin{center}
\includegraphics[width=15cm]{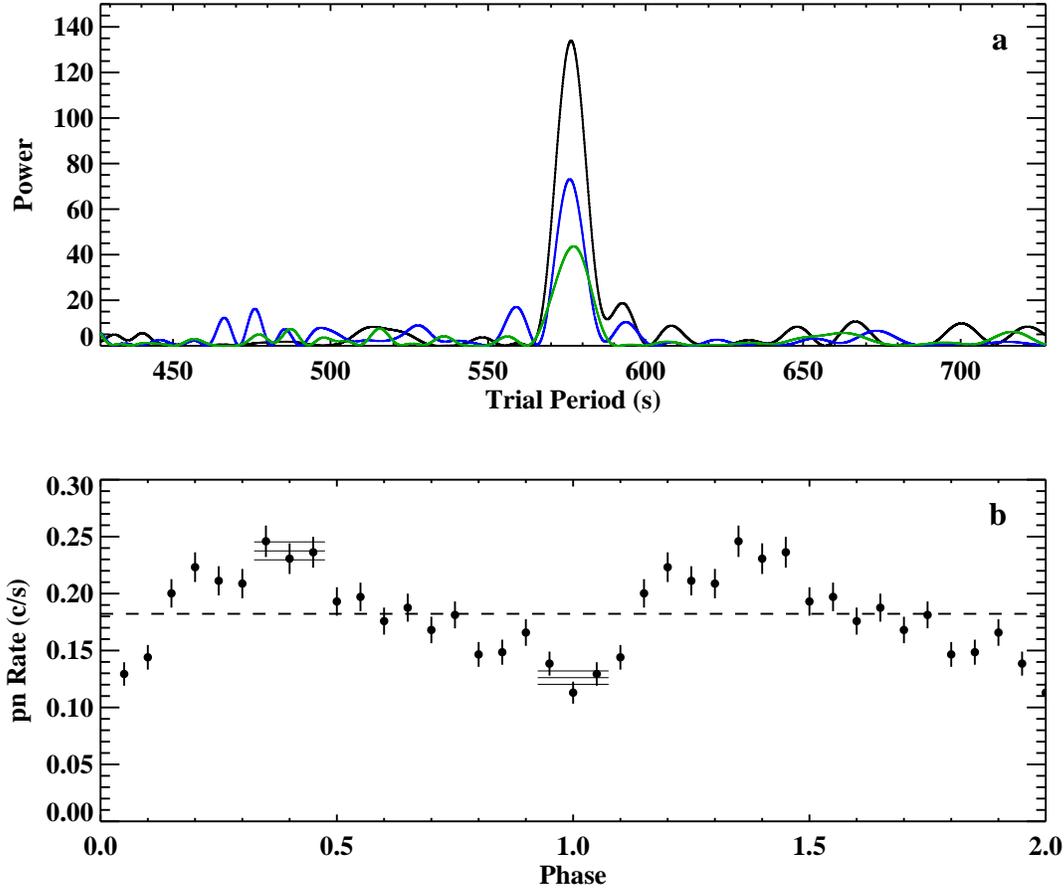}
\caption{\small {\em (a)} Periodograms for 0.3--12\,keV light curves from {\em XMM}
pn (black), MOS1 (blue), and MOS2 (green), showing the detection of a signal at
$576.3\pm 0.6$\,s.  For the periodogram, the power is calculated using the 
$Z_{1}^{2}$ test.  {\em (b)} Folded pn light curve in the 0.3--12\,keV band.  The
horizontal solid lines indicate the phases used to determine the maximum and minimum 
rates.  The difference between the maximum and minimum count rates divided by the 
sum of the maximum and minimum rates is the pulsed fraction.\label{fig:period}}
\end{center}
\end{figure*}

We produced 0.3--12\,keV event lists for the three {\em XMM} instruments, including
only the photons in a $40^{\prime\prime}$ circular region centered on the position 
of IGR~J14091--6108.  The times of the individual events were shifted to the
solar system barycenter using the SAS tool {\ttfamily barycen}.  We used the 
$Z_{1}^{2}$ (Rayleigh) test \citep{buccheri83} to make periodograms from the event 
list.  We searched for a periodic signal in the pn data by making a periodogram 
with 20000 time bins between 1 and 2000\,s.  One significant peak is present 
with a peak value of $S = 134$ (see Figure~\ref{fig:period}a).  The false alarm 
probability (FAP) is given by 0.5\,$e^{-S/2}$ multiplied by the number of trials, 
corresponding to a FAP of $8.5\times 10^{-26}$.  The period is $576.3\pm 0.6$\,s, 
where the 1-$\sigma$ errors are given by the periods where the periodogram has 
fallen to $S$--1.  The MOS1 and MOS2 instruments showed significant peaks at
$575.9\pm 0.8$\,s and $577.1\pm 1.3$\,s, respectively, which are both consistent
with the period measured by pn (Figure~\ref{fig:period}a).  We produced a folded 
light curve, which is shown in Figure~\ref{fig:period}b.  An epoch of zero phase 
(minimum in the folded light curve) is MJD TDB $57224.0971\pm 0.0003$.  To align
the {\em XMM} and {\em NuSTAR} phases, we barycentered the {\em NuSTAR} event
lists and folded the {\em NuSTAR} photons on the {\em XMM} ephemeris.

\begin{figure*}
\begin{center}
\includegraphics[width=15cm]{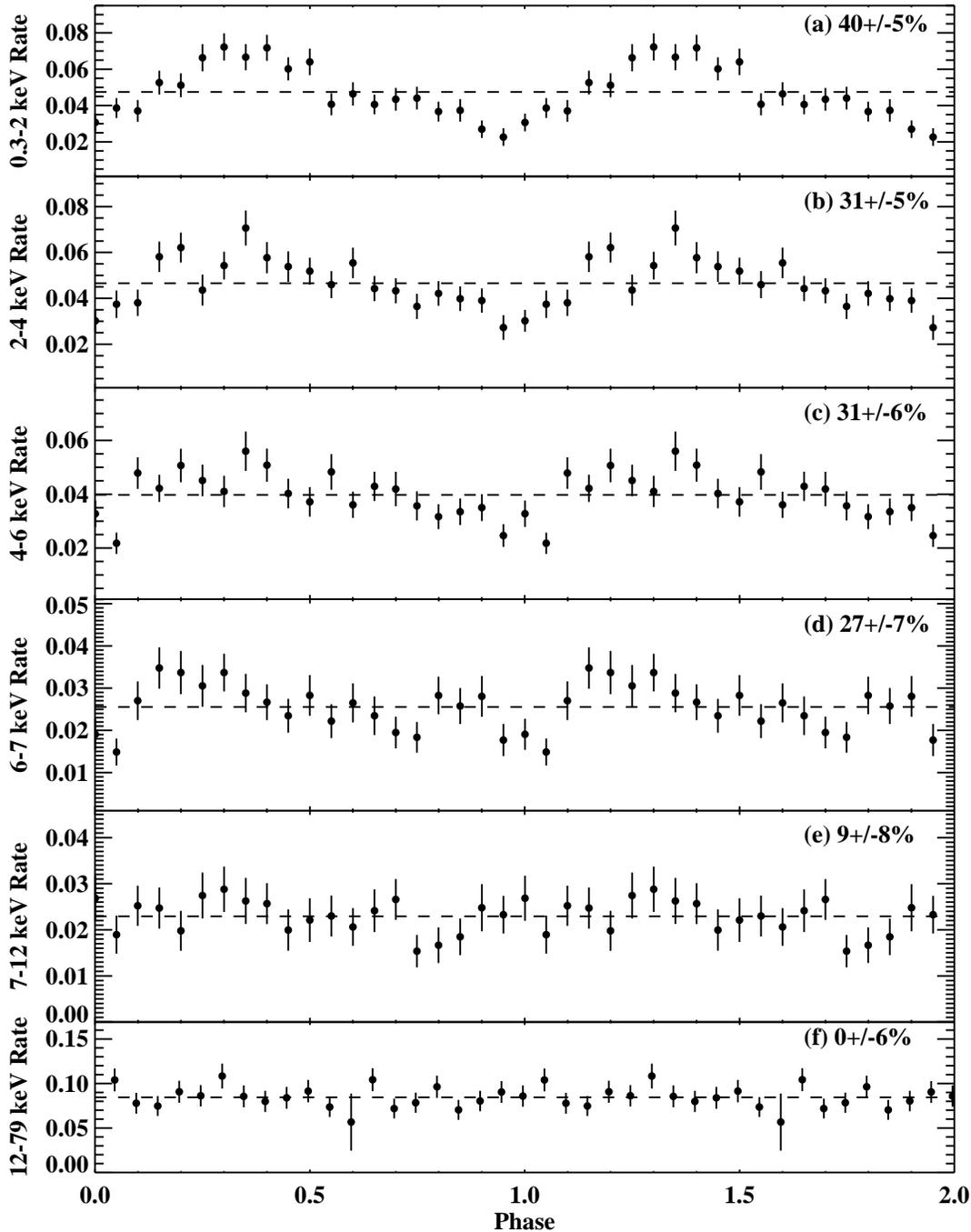}
\caption{\small The X-ray light curves folded on the 576.3\,s period.
Panels {\em (a)-(e)} show folded light curves for five {\em XMM} pn 
energy bands, and panel {\em (f)} shows the 12--79\,keV band from 
{\em NuSTAR}.  Each panel shows the calculated value of the pulsed 
fraction with its 1-$\sigma$ uncertainty.  The same values are 
plotted vs. energy in Figure~\ref{fig:amp}.\label{fig:folded_energy}}
\end{center}
\end{figure*}

\begin{figure*}
\begin{center}
\includegraphics[width=10cm]{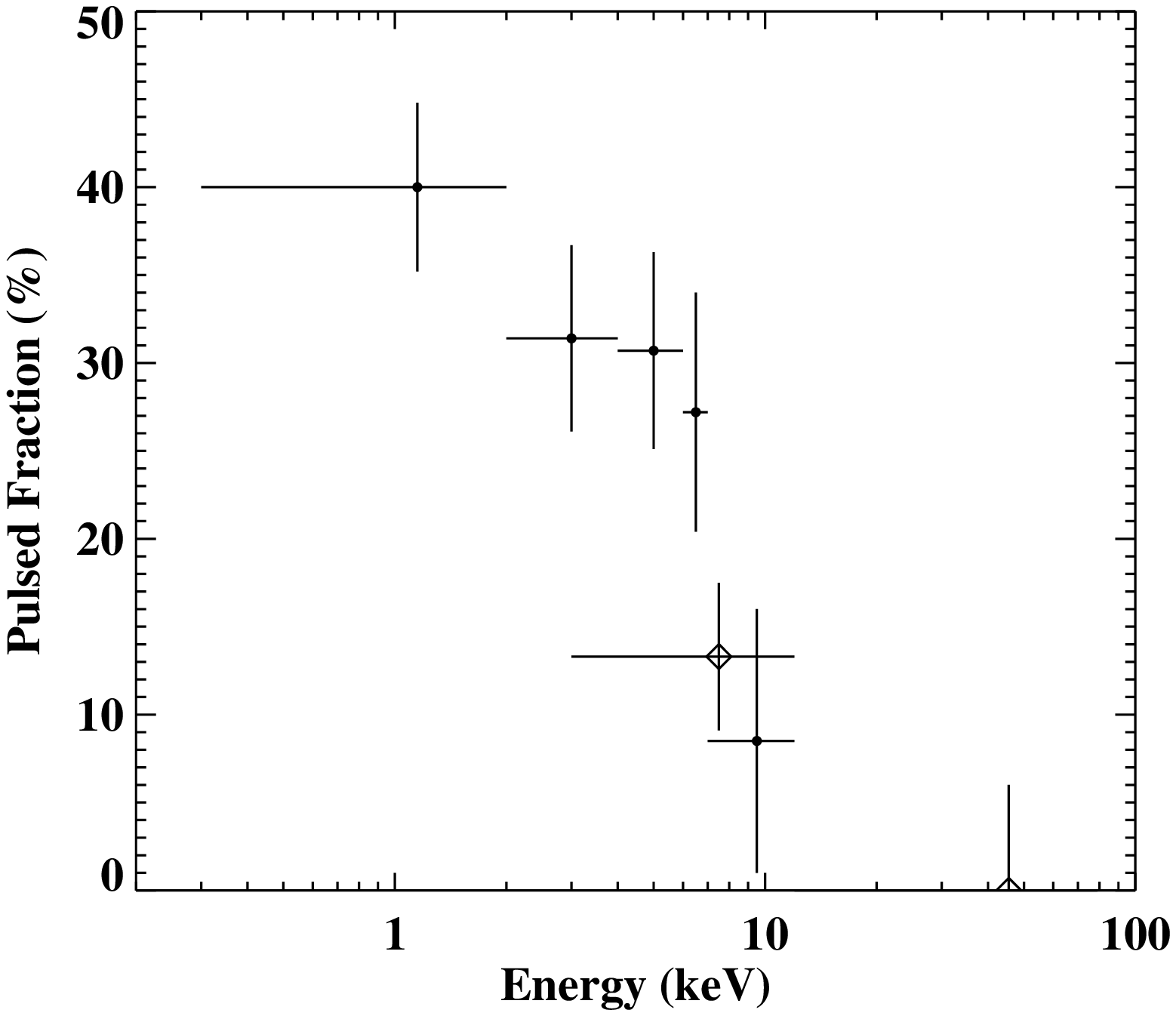}
\caption{\small The pulsed fraction of the 576.3\,s period versus energy.  
The filled circles mark the {\em XMM} pn measurements, and the diamonds 
mark the {\em NuSTAR} (FPMA+FPMB) measurements.  The errors shown are 
1-$\sigma$.\label{fig:amp}}
\end{center}
\end{figure*}

We calculated the pulsed fraction of the signal by defining maximum and minimum 
phase ranges using the 0.3--12\,keV folded light curve (Figure~\ref{fig:period}b).  
The pulsed fraction is defined as the absolute value of the difference between 
the maximum and minimum count rate divided by the sum of these quantities.  In 
the 0.3--12\,keV range, it is $30\pm 3$\%.  The folded light curves from 0.3 to 
79\,keV are shown in Figure~\ref{fig:folded_energy}, and the values for the
pulsed fraction are indicated on Figure~\ref{fig:folded_energy} and plotted 
in Figure~\ref{fig:amp}.  The {\em XMM} measurements show a strong decrease 
in the pulsed fraction with energy with the largest drop being at 7\,keV.  The 
{\em NuSTAR} measurements also show a drop with energy, with a pulsed fraction 
of $13\pm 4$\% in the 3--12\,keV band and $0\pm 6$\% (no detection) in the 
12--79\,keV band.  

A period of 576\,s is fairly typical for the spin period of a white dwarf in 
a CV of Intermediate Polar (IP) type.  In the 2014 update of the \cite{rk03} 
CV catalog\footnote{see http://vizier.u-strasbg.fr/viz-bin/VizieR-3?-source=B/cb/cbdata}, 
there are 78 IP-type CVs (CV/IPs) with white dwarf spin periods between 128 and 12071\,s, 
and 20 systems have periods faster than 576\,s.  The shortest CV/IP orbital period
is 1552\,s, and the median orbital period is 4.1\,hr.  Thus, if IGR~J14091--6108
is a CV/IP, the detected period is much more likely to be the white dwarf spin
period, and the orbital period is probably between several and $\sim$100 times
longer \citep{hong12}

\subsection{Optical Spectra}

\begin{figure*}
\begin{center}
\includegraphics[width=10cm,angle=270]{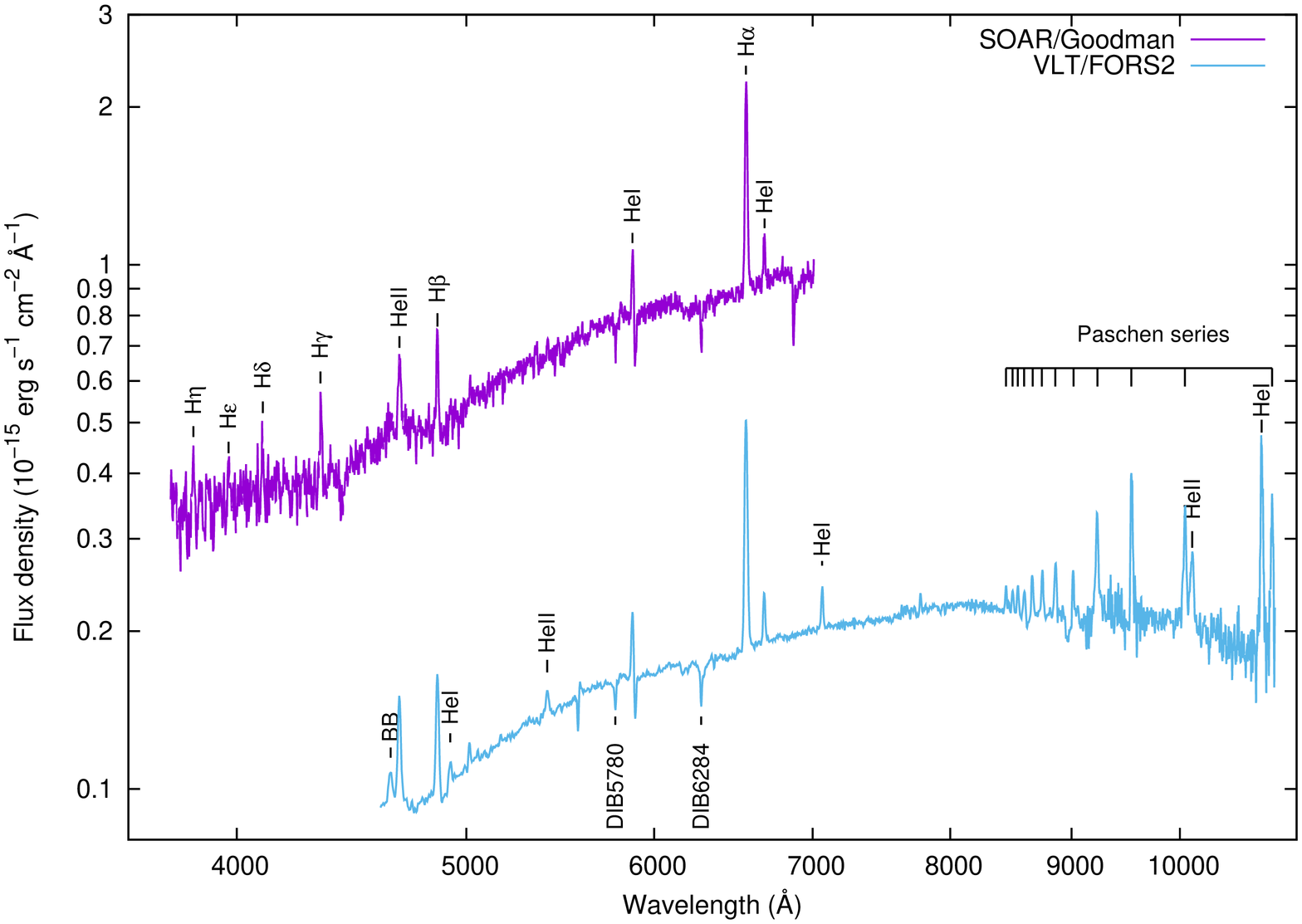}
\caption{\small Two optical spectra of the IGR~J14091--6108 counterpart
(VVV~J140845.99--610754.1).  The purple spectrum was taken at SOAR in
2015 January with the Goodman spectrograph, and the light blue spectrum 
was taken in 2015 April at VLT with the FORS2 spectrograph.  The 
emission lines from the source, and the lines due to interstellar 
absorption are labeled.  See Table~\ref{tab:speclines} for detailed 
line information and parameters.\label{fig:optical1}}
\end{center}
\end{figure*}

Figure~\ref{fig:optical1} displays the flux-calibrated SOAR/Goodman
(magenta) and VLT/FORS2 (blue) spectra of IGR~J14091$-$6108, on which 
all the detected spectroscopic features are marked. Likewise, 
Table~\ref{tab:speclines} lists their main parameters, i.e. their 
central wavelengths ($\lambda_{\rm c}$), equivalent widths ($\mathring{W}$), 
full-widths at half-maximum (FWHM), and intrinsic fluxes ($F_{\rm Line}$), 
obtained through single-Gaussian fitting. The FWHMs were quadratically 
corrected for the instrumental broadening and the underlying continua 
were locally assessed with a first-order polynomial. The continuum level 
being the primary source of inaccuracy, each measurement was repeated 
several times with different placements within the same wavelength range 
to obtain a set of values that eventually averaged out. The listed 
uncertainties are therefore the scatter to the mean rather than just 
statistical. 

The optical spectrum is very rich, with a wealth of spectral features
that includes the Balmer and Paschen series as well as several
signatures of \ion{He}{1} and \ion{He}{2}, all in emission. The Bowen
Blend at 4640~\AA, typical of irradiated accretion disks and/or
companion stars in accreting systems, is also clearly observed with
FORS2 and perhaps with Goodman. The spectrum also appears to be
strongly variable, the continuum and emission lines being roughly
fives and three times brighter in the Goodman spectrum, respectively.

Besides features intrinsic to IGR~J14091$-$6108, we also report two
diffuse interstellar bands (DIBs) centered at 5780~\AA\ and
6284~\AA. DIBs are strongly correlated to the ISM extinction along the
line-of-sight of the sources in which they are detected and we can
assess the latter using the relationships between their equivalent
widths and $E(B-V)$ given in \citet{1994Jenniskens}. Nonetheless,
DIB6284 is likely contaminated with some atmospheric absorption
troughs and we therefore rely on DIB5780 only, for which
\citet{1994Jenniskens} obtain
$\frac{\mathring{W}}{E(B-V)}\,=\,0.647\pm0.053$. We measure 
$\mathring{W}_{\rm 5780}\,=\,0.97\pm0.11$ and 
$\mathring{W}_{\rm 5780}\,=\,1.11\pm0.07$ in the Goodman and FORS2 
spectra, respectively, which, once averaged out, leads to
$E(B-V)\,=\,1.615\pm0.166$. Using the relationship 
$A_{V}\,=\,R_{V}\times E(B-V)$ and an average total-to-selective
extinction ratio $R_{V}=3.1$, we thus derive 
$A_{V}\,=\,5.01\pm0.51$, which is roughly consistent with a 
distance between 3 and 4 kpc based on the 3D Galactic extinction 
map derived in \citet{marshall06} for a line-of-sight 9\,\farcs4 away 
from that of IGR~J14091$-$6108.

\begin{figure*}
\begin{center}
\includegraphics[width=10cm,angle=270]{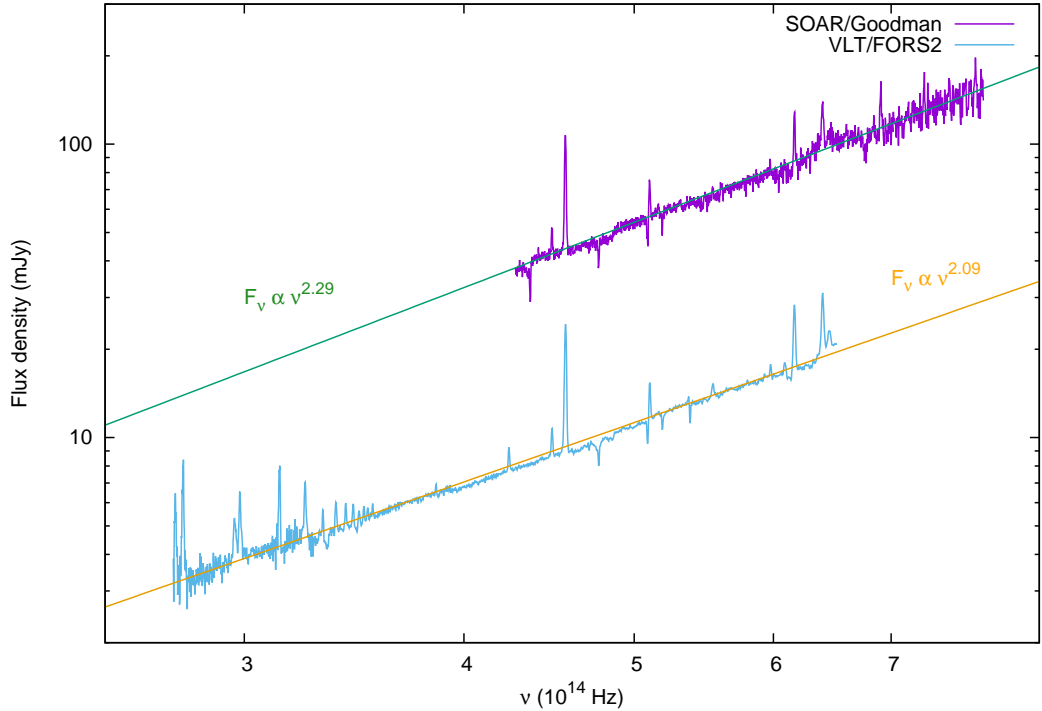}
\caption{\small The same SOAR and VLT optical spectra of IGR~J14091--6108
as shown in Figure~\ref{fig:optical1}.  The spectra are dereddened, shown
as a function of frequency, and the continua are each compared to a power-law
function.\label{fig:optical2}}
\end{center}
\end{figure*}

With the knowledge of $A_{V}$, it is now possible to correct 
IGR~J14091$-$6108 emission from the ISM extinction and
Figure~\ref{fig:optical2} displays the extinction-corrected Goodman and
FORS2 spectra expressed in Hz vs mJy. The first result is that both
continua are roughly best-fit with power laws with 2.29 and 2.09
spectral indices, respectively, typical of the Raleigh-Jeans tail of a
black body emitter. We can also estimate the Balmer decrements and we
find H{\small $\alpha$}/H{\small $\beta$} = $1.03\pm0.20$ from the
FORS2 spectrum as well as
H{\small $\alpha$}/H{\small $\beta$} = $1.20\pm0.23$, H{\small
$\gamma$}/H{\small $\beta$} = $1.07\pm0.26$, H{\small
$\delta$}/H{\small $\beta$} = $1.03\pm0.27$, H{\small
$\epsilon$}/H{\small $\beta$} = $0.86\pm0.27$, and H{\small
$\eta$}/H{\small $\beta$} = $1.38\pm0.32$ from the
Goodman spectrum. It is clear that all the
values are consistent with unity, which is typical of
Cataclysmic Variables (CVs) and some microquasars 
\citep[see, e.g.,][]{1980Williams, 1988Williams, 2014Rahoui}. Based on the
wealth of emission lines, spectral variability, Raleigh-Jeans
continuum, and flat Balmer decrements around unity, it is thus very
likely that the optical emission of IGR~J14091$-$6108 originates in an
optically-thick irradiated accretion disk in a CV.

\subsection{X-ray Spectrum}

For spectral analysis, we jointly fitted the {\em XMM} (pn, MOS1, and MOS2) and 
{\em NuSTAR} (FPMA and FPMB) data in the 0.3--79\,keV energy range.  We used the
XSPEC package \citep{arnaud96} and performed the fitting with $\chi^{2}$ minimization.
We fitted with an absorbed power-law model, using \cite{wam00} abundances and 
\cite{vern96} cross sections for the absorption.  We also allowed for normalization 
differences between instruments by introducing a multiplicative constant.  Thus, the 
overall model in XSPEC notation was {\ttfamily constant*tbabs*pegpwrlw}.  This showed 
that the spectrum is hard with $\Gamma\sim 0.8$, but the fit was poor with 
$\chi^{2} = 1242$ for 346 degrees of freedom (dof).  The largest residuals 
appear in the iron line region, indicating a strong emission line.  Adding a
Gaussian with $E_{\rm line} = 6.56\pm 0.04$\,keV and $\sigma_{\rm line} = 0.35\pm 0.05$\,keV 
provides a large improvement in the fit (to $\chi^{2}/\nu = 711/343$), but negative
residuals in the high energy part of the spectrum indicate curvature in the spectrum
and possibly a cutoff.

We changed the continuum model from a power-law to a thermal Bremsstrahlung component
to allow for curvature in the model and also because this is the emission mechanism
that is thought to operate in CV/IPs.  While the model 
{\ttfamily constant*tbabs*(gaussian+bremss)} does not provide a good fit to the data 
($\chi^{2}/\nu = 997/343$), partial covering absorption is typically used when fitting 
CV/IPs \citep{srr05,mukai15}, and, in our case, using {\ttfamily pcfabs} with 
$N_{\rm H} = (1.4\pm 0.3)\times 10^{23}$\,cm$^{-2}$ and a covering fraction of 
$0.68^{+0.02}_{-0.03}$ improves the fit greatly to $\chi^{2}/\nu = 516/341$.  
With this model, the Bremsstrahlung temperature is very high, $kT > 167$\,keV
(90\% confidence limit), but there are still residuals at the high energy end
that we suspect are related to the presence of a reflection component. 

\begin{figure*}
\begin{center}
\includegraphics[width=15cm,angle=0]{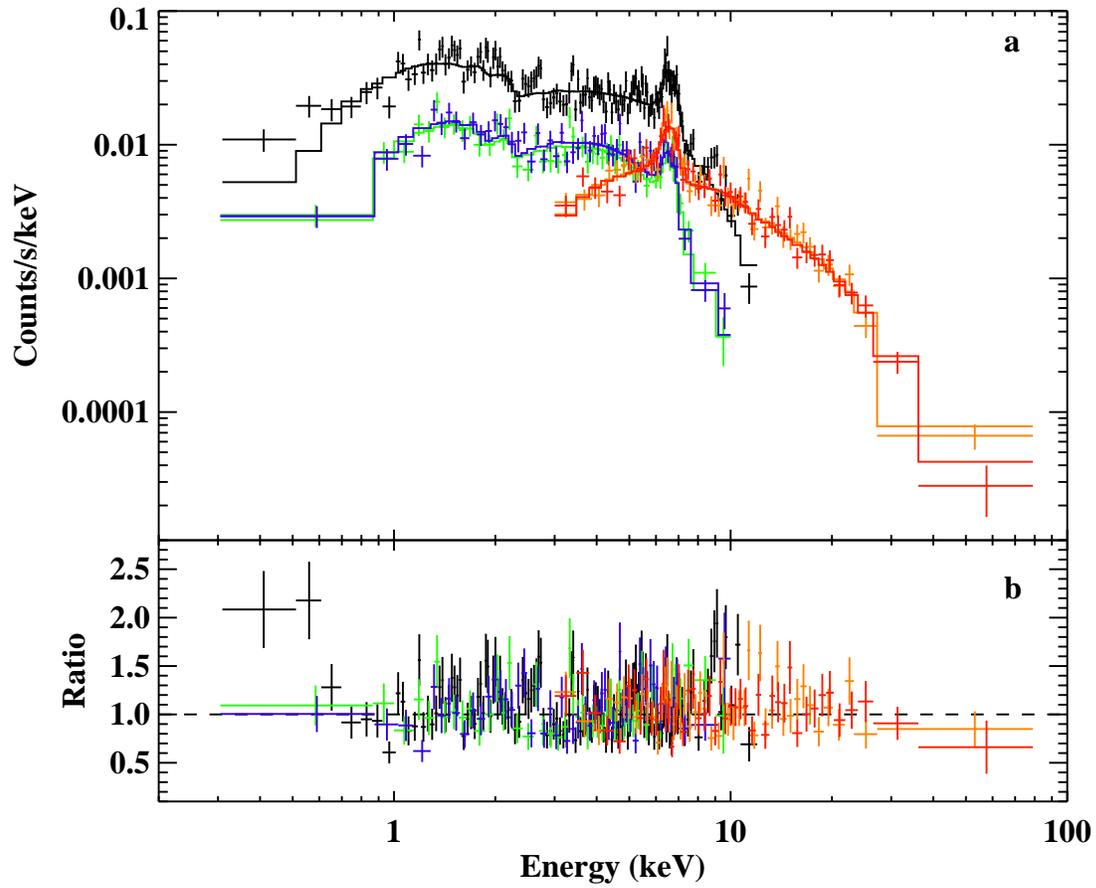}
\caption{\small {\em (a)} {\em XMM} and {\em NuSTAR} energy spectrum (folded 
through the instrument response) fitted with a model consisting of 
{\ttfamily constant*tbabs*pcfabs*(gaussian+reflect*bremss)}.  The black, 
light green, and blue spectra are for the {\em XMM} pn, MOS1, and MOS2 
instruments, respectively.  The orange and red spectra are for 
{\em NuSTAR} FPMA and FPMB, respectively. {\em (b)} The residuals in 
terms of the data-to-model ratio.\label{fig:spectrum_counts}}
\end{center}
\end{figure*}

Reflection of the hard X-ray emission off the white dwarf surface is often 
included in models when fitting CV/IP spectra, and recent work with {\em XMM}
and {\em NuSTAR} spectra show strong evidence for this component in three other
CV/IPs \citep{mukai15}.  For IGR~J14091--6108, we added a reflection component
using the {\ttfamily reflect} model in XSPEC.  This model is based on
\cite{mz95}, which is for reflection of direct emission from neutral material
and includes the dependence on viewing angle.  By convolving {\ttfamily bremss}
with {\ttfamily reflect}, the model includes both a direct and a reflected
Bremsstrahlung component where the strength of the reflected component depends
on the amplitude parameter, $\Omega/2\pi$.  In our case, if the reflection 
amplitude is left as a free parameter, it will increase to values above 1.0.
However, assuming that we see 100\% of the direct emission, values of 
$\Omega/2\pi$ above unity are not physically possible for reflection from the 
white dwarf surface.  Thus, we fix the reflection amplitude to 1.0, resulting
in a fit with $\chi^{2}/\nu = 472/339$.  The spectrum fit with this model is
shown in Figure~\ref{fig:spectrum_counts}, and the parameters are given in
Table~\ref{tab:spec1}.  Although the Bremsstrahlung temperature drops when
the reflection component is added, we still measure $kT = 81^{+31}_{-20}$\,keV, 
which, as we discuss in Section 4, is high for a CV/IP.  For this fit, the 
0.3--79\,keV unabsorbed (i.e., with the interstellar but not the partial covering 
column density set to zero) flux is $1.1\times 10^{-11}$\,erg\,cm$^{-2}$\,s$^{-1}$.

While the actual reflection component has three main features: the Compton hump
above 10\,keV, the iron edge at 7.1\,keV (for neutral iron), and an iron 
fluorescence line at 6.4\,keV (also for neutral iron), the \cite{mz95} model
only includes the first two features, which is one reason that we include the 
Gaussian in the model above.  If the emission line was due only to reflection
from the white dwarf surface, we would expect a relatively narrow line centered 
at 6.4\,keV; however, as shown in Table~\ref{tab:spec1}, we measure a broad 
line with $\sigma_{\rm line} = 0.28\pm 0.04$\,keV at an energy above 6.4\,keV 
($E_{\rm line} = 6.59\pm 0.04$\,keV).  It is very likely that this is due to
contributions from higher ionization states coming from the hot material in
the accretion column, and this is common for CV/IPs \citep{hm04}.  We modified 
the model to include two lines with energies fixed to 6.4\,keV (neutral iron) 
and 6.7\,keV (He-like iron).  This more physical representation of the emission 
lines provides an equivalently good fit ($\chi^{2}/\nu = 473/339$), but the 
lines are still relatively broad ($\sigma_{\rm line} = 0.24\pm 0.05$\,keV).  
Adding a third line at 6.97\,keV (H-like iron) improves the fit 
($\chi^{2}/\nu = 458/338$), and these lines are significantly narrower 
($\sigma_{\rm line} = 0.07\pm 0.04$\,keV), which is consistent with line widths 
measured for other CV/IPs \citep{hm04}.  The line normalizations in 
Table~\ref{tab:spec1} include 90\% confidence uncertainties, suggesting
that all the lines are significantly detected.  Thus, the three-line model is 
very likely the correct interpretation based on the match between measured and 
expected line widths as well as on statistical grounds.  In this model, the 
equivalent widths of the three lines are $320\pm 60$, $160\pm 40$, and 
$170\pm 50$\,eV for neutral, He-like, and H-like iron, respectively.

\begin{figure*}
\begin{center}
\includegraphics[width=15cm,angle=0]{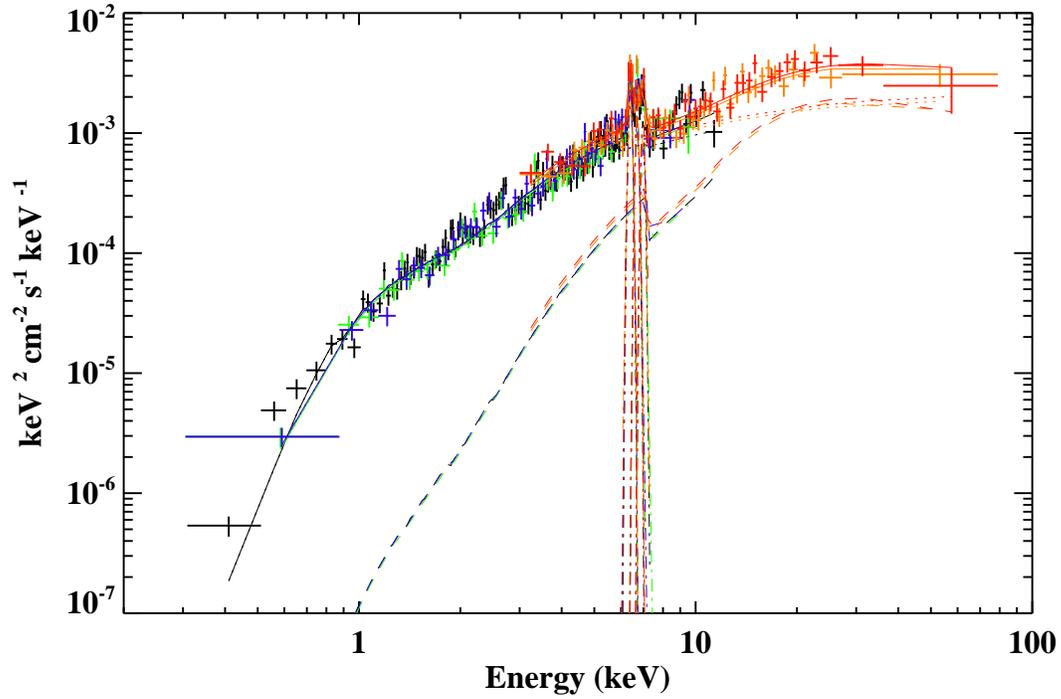}
\caption{\small Unfolded {\em XMM} and {\em NuSTAR} energy spectrum fitted with 
the IP Mass model fit shown in Table~\ref{tab:spec2}.  The components are 3 
Gaussians {\em (dash-dotted lines)}, reflection {\em (dotted line)}, and the 
IP Mass model {\em (dashed)}.  The spectra are plotted using the same colors as 
for Figure~\ref{fig:spectrum_counts}.\label{fig:spectrum_efe}}
\end{center}
\end{figure*}

To explore the possible physical cause of the high temperature derived from the
fitting the continuum with a Bremsstrahlung model, we replaced the Bremsstrahlung
component with the IP mass (IPM) model of \cite{srr05}.  For the IPM model, the
hardness and cutoff energy of the predicted spectrum are both set by the white
dwarf mass, and the only two free parameters are $M_{\rm WD}$ and the normalization.
Physically, this connection between the $M_{\rm WD}$ and the spectrum is related
to the maximum shock temperature in the accretion column (Suleimanov et al. 
2005\nocite{srr05}; Hailey et al., submitted to ApJ).  The emission mechanism 
is still Bremsstrahlung, but the model includes a range of temperatures.  The
results for a model including direct and reflected IPM components, partial
covering, and three Gaussians are reported in Table~\ref{tab:spec2}.  As before, 
we fix the reflection amplitude to 1.0, and the other reflection parameters are 
consistent to the values found in the Bremsstrahlung fits.  Also, there is little 
or no change in the continuum absorption and the emission line parameters.  
The IPM model parameters imply a high white dwarf mass, $M_{\rm WD} > 1.38$\Msun.
The components of the spectrum are shown in Figure~\ref{fig:spectrum_efe}.

We note that none of the models we have used provide formally acceptable 
fits, with the best one having a reduced-$\chi^{2}$ of 1.36 for 338 dof.
\cite{mukai15} obtained similar fit qualities for his {\em XMM}+{\em NuSTAR}
spectra of CV/IPs and suggested that it is related to calibration differences
between instruments, and this is also a possibility in our case.  However, 
we explored some possibilities for improving the fits.  For example, the 
fit improves if we start with the model given in Table~\ref{tab:spec2} and 
allow $\Omega/2\pi$ to increase above 1.0.  Although likely unphysical, the 
fit improves from a reduced-$\chi^{2}$ of 1.42 (for 338 dof) to 1.33 (for 337 dof)
at $\Omega/2\pi = 5.0$.  Also, allowing the reflection component to dominate at
high energies causes the IPM component to soften, and, for the extreme 
assumption that $\Omega/2\pi = 5.0$, $M_{\rm WD}$ drops to 1.2\Msun.  The
fit can also be improved by modifying the modeling of the absorption.  In 
some CV/IPs, there is evidence for complex absorption patterns, and other
authors have modeled this with two partial coverers \citep[e.g.,][]{boh00}.
Adding a second partial coverer to the 3 Gaussian Bremsstrahlung model in 
Table~\ref{tab:spec1} improves the reduced-$\chi^{2}$ from 1.36 (for 338 dof)
to 1.23 (for 336 dof).  Although this is a significant improvement, this
model leads to an interstellar column density of 
$N_{\rm H} = (0.16\pm 0.10)\times 10^{22}$\,cm$^{-2}$, which is much lower than 
the value implied by our measured optical extinction.  Using the relation 
from \cite{go09}, an $A_{V}$ of $5.0\pm 0.5$ magnitudes corresponds to 
$N_{\rm H} = (1.1\pm 0.1)\times 10^{22}$\,cm$^{-2}$.  Thus, while both high
reflection amplitude and additional partial covering absorption lead to 
improved fits, the former is not consistent with the physical scenario 
and the latter leads to unrealistically low values for the interstellar 
column density.

To avoid the complexities of the low energy part of the spectrum, we
also explored fitting just the {\em NuSTAR} spectrum above 10\,keV.  
Fitting with a Bremsstrahlung model gives a reduced-$\chi^{2}$ of 1.62 
for 37 dof, and the 90\% confidence lower limit on the temperature 
is 112\,keV.  Adding a reflection component with $\Omega/2\pi = 1$, which 
makes the model {\ttfamily constant*reflect*bremss}, provides a good fit 
(reduced-$\chi^{2}$ = 1.14 for 35 dof), and the Bremsstrahlung temperature 
is $66^{+33}_{-22}$\,keV.  With {\ttfamily constant*reflect*ipm} and 
$\Omega/2\pi = 1$, we find $M_{\rm WD} > 1.31$\Msun, which is consistent 
with the high mass values we obtain when fitting the entire spectrum.

\section{Discussion}

The results indicate that we can identify IGR~J14091--6108 as a CV/IP 
with a high degree of confidence.  The $576.3\pm 0.6$\,s periodicity 
shows that there is a magnetized compact object in the system, and 
the hardness of the X-ray spectrum is typical of only a CV/IP or an
accreting pulsar in an HMXB.  The optical spectrum is dominated by an 
optically-thick irradiated accretion disk, and no evidence for emission 
from the companion star is found, indicating that it must be a low-mass 
star and ruling out the HMXB possibility. There is also strong long-term 
optical variability in the continuum and emission lines, the origin of 
which is not clear.  IPs are known to be strongly variable in the optical 
regime and often show large spectroscopic modulations associated with the 
spin period of the white dwarf and/or the orbital period 
\citep[see e.g.][]{still98,belle03,scaringi11}. While this may 
be the case for IGR~J14091$-$6108, especially for the spectral lines, 
we believe that the large brightness difference between the FORS2 and 
Goodman optical spectra is mainly due to different accretion rates 
within the accretion disk.

The X-ray spectrum is also consistent with a CV/IP, and we confirm
the presence of the strong iron line originally reported by
\cite{tomsick16} using {\em Chandra} measurements.  With {\em XMM}, 
we find that the line emission is consistent with being a combination 
of lines from three ionization states with widths of $0.07\pm 0.04$\,keV
and equivalent widths in the 160--320\,eV range.  These values are
typical of iron line complexes seen in CV/IPs \citep{hm04}.  Also, 
combining the 0.3--79\,keV flux of $1.1\times 10^{-11}$\,erg\,cm$^{-2}$\,s$^{-1}$
with our estimated distance of 3--4\,kpc from the optical extinction, 
we obtain an X-ray luminosity of 
$1.6\times 10^{34}$\,$d_{3.5\,\rm {kpc}}^{2}$\,erg\,s$^{-1}$, 
which is consistent with expectations for CV/IPs.  In the \cite{bird16} 
catalog, there are 40 confirmed and seven candidate CV/IPs.  Of the 
confirmed systems, 18 have IGR names, making IGR~J14091--6108 the 
19th confirmed CV/IP discovered by {\em INTEGRAL} \citep{bird16}.
However, we note that IGR~J14091--6108 itself is not in the \cite{bird16}
catalog, suggesting the possible presence of a much larger number
of CV/IPs close to the {\em INTEGRAL} detection limit.

The strong drop in the pulsed fraction with energy that we see for
IGR~J14091--6108 has been seen for other CV/IPs.  \cite{taylor97}
report this energy dependence for the CV/IPs AO~Psc and V1223~Sgr.
For V1223~Sgr, phase-resolved spectroscopy shows that the energy 
dependence is primarily related to changes in the column density of
absorbing material local to the CV \citep{hayashi11}, and we suspect
that this is also the case for IGR~J14091--6108.

While IGR~J14091--6108 is typical of CV/IPs in many respects, it is
a candidate system for having a higher than typical white dwarf mass.
Using the IP Mass model that \cite{srr05} applied to 14 CV/IPs, 
we obtain a mass in excess of 1.3\Msun~for IGR~J14091--6108, while 
\cite{srr05} find masses between $0.50\pm 0.05$\Msun~(for EX Hya) and 
$1.00\pm 0.20$\Msun~(for V1062 Tau).  \cite{yuasa10} also used spectral
fitting to estimate white dwarf masses for 17 CV/IPs, and they found
masses as high as $\sim$1.2\Msun~for V709~Cas, PQ~Gem, and NY~Lup.  
Hailey et al. (submitted) fit spectra from several X-ray satellites
including {\em NuSTAR} for the CV/IP IGR~J17303--0601, which has a 
relatively massive white dwarf.  With the {\em NuSTAR} spectrum, partial
covering absorption, and reflection, they measured a white dwarf mass
of $1.16\pm 0.12$\Msun.  If they replaced the IPM model with a Bremsstrahlung
model, they obtained a temperature of $34\pm 2$\,keV.  From this comparison, 
the temperature of $81^{+31}_{-20}$\,keV (or $66^{+33}_{-22}$\,keV for the 
$>$10\,keV fit) that we measure for IGR~J14091--6108 is suggestive of 
mass in excess of $\sim$1.2\Msun.  Hailey et al. discuss the CV/IP spectra 
in the context of the diffuse hard X-ray emission from the Galactic center 
\citep{perez15}, which may be caused by an unresolved population of magnetic 
CVs with white dwarfs that are more massive than the average Galactic 
population \citep{krivonos07_cv}.

One reason for interest in massive white dwarfs is the question of whether 
the progenitors of type Ia supernovae (SNe) are merging white dwarfs or 
accreting white dwarfs that detonate when they reach the Chandrasekhar 
limit. Based on the relatively low soft X-ray luminosities from nearby 
elliptical galaxies and galaxy bulges, \cite{gb10} and \cite{distefano10a} 
argue that $<$5\% of type Ia SNe are from accreting white dwarfs 
\citep[although subsequent works showed that the X-rays could be attenuated by stellar winds, accretion winds, or white dwarf atmospheres;][]{distefano10b,nielsen13a}.  Along 
these same lines, pre-explosion imaging of individual nearby type Ia SNe 
place interesting constraints on accreting and nuclear-burning white dwarfs 
as the SN progenitors \citep{liu12,nielsen12,nielsen13b,nielsen14,graur14}.  
On the other hand, the white dwarfs in CVs are, on average, more massive 
than white dwarfs that have not undergone mass accretion from a companion 
\citep{zsg11}, implying that accretion leads to a significant increase in
the masses of the white dwarfs.  Given the possibility that IGR J14091--6108 
may harbor a white dwarf with a mass very close to the Chandrasekhar limit, 
follow-up observations to measure $M_{\rm WD}$ using other techniques should 
be high priority. Based on the 576\,s spin period, an orbital period of 
several hours would be expected, and optical photometry might be used to 
search for this period. It may then be possible to obtain a radial velocity 
curve, but this may need to be done in the near-IR where we might see
absorption lines from the companion star's photosphere.


\begin{table*}
\caption{Observations of IGR~J14091--6108\label{tab:obs}}
\begin{minipage}{\linewidth}
\begin{tabular}{cccccccc} \hline \hline
Observatory  & ObsID         & Instrument & Start Time (UT) & End Time (UT) & Exposure Time (ks)\\ \hline\hline
{\em XMM}    & 0761940301    & pn         & 2015 July 21, 2.22 h  & 2015 July 21, 10.03 h & 24.7\\
''           & ''            & MOS1       & 2015 July 21, 1.89 h  & 2015 July 21, 10.15 h & 28.9\\
''           & ''            & MOS2       & ''                    & ''                    & ''\\
{\em NuSTAR} & 30101001002   & FPMA       & 2015 July 20, 21.85 h & 2015 July 21, 10.52 h & 22.9\\
''           & ''            & FPMB       & ''                    & ''                    & ''\\
VLT          & 095.D-0972(A) & FORS2      & 2015 April 13, 6.64 h   & 2015 April 13, 7.49 h & 2.2\\
SOAR         & ---           & Goodman    & 2015 January 15, 8.45 h & 2015 January 15, 8.62 h & 0.60\\
             &               & Spectrograph &                     &                       &     \\ \hline
\end{tabular}
\end{minipage}
\end{table*}

\begin{table*}
\caption{Optical lines in the IGR~J14091$-$6108 FORS2 and Goodman spectra.\label{tab:speclines}}
\begin{minipage}{\linewidth}
\begin{tabular}{l|c|ccc||ccc} \hline\hline
&&\multicolumn{3}{c||}{FORS2}&\multicolumn{3}{c}{Goodman}\\
\hline
Element&${\lambda_{\rm c}}$\footnote{Measured wavelength in \AA}&${\mathring{W}}$\footnote{Equivalent widths in \AA}&FWHM\footnote{Full-width at half-maximum in \kms, quadratically corrected for instrumental broadening}&${F_{\rm line}}$\footnote{Intrinsic line flux in units of $10^{-15}\,\,{\rm erg\,\,cm}^{-2}\,\,{\rm s}^{-1}$}&${\mathring{W}}$&FWHM&${F_{\rm line}}$\\
\hline
H{$\eta$}	&	3835	&--&--&--&$-3.4\pm0.7$&$637\pm96$&$1.10\pm0.20$\\
H{$\epsilon$}	&	3968	&--&--&--&$-2.4\pm1.0$&$579\pm307$&$0.84\pm0.25$\\
H{$\delta$}	&	4101	&--&--&--&$-2.3\pm0.6$&$281\pm78$&$0.83\pm0.20$\\	
H{$\gamma$}	&	4341	&--&--&--&$-5.4\pm1.2$&$690\pm145$&$2.00\pm0.37$\\
BB\footnote{The Bowen Blend, which is the \ion{C}{3}+\ion{N}{3} complex.} &	4645	&--3.2$\pm$1.1	     &1090$\pm$135	&0.31$\pm$0.12 &--&--&--\\
\ion{He}{2}	&	4684	&--9.7$\pm$2.8	     &666$\pm$72 	&0.93$\pm$0.23&$-5.9\pm0.9$&$856\pm95$&$2.78\pm0.38$\\
H{$\beta$}	&	4861	&--10.1$\pm$0.8          &539$\pm$51		&1.02$\pm$0.07&$-5.4\pm1.1$&$494\pm43$&$2.94\pm0.27$\\
\ion{He}{1}    	&	4922	&--1.1$\pm$0.3	     &523$\pm$48		&0.12$\pm$0.04 &--&--&--\\
\ion{He}{2}     	&	5409	&-1.6$\pm$0.4            &551$\pm$72 		&0.25$\pm$0.04 &--&--&--\\
\ion{He}{1} 	&	5875	&--4.9$\pm$0.9	     &353$\pm$57		&0.83$\pm$0.05&$-4.0\pm0.5$&$492\pm82$&$3.21\pm0.44$\\
H{$\alpha$}	&	6561	&--37.6$\pm$1.5          &652$\pm$45 		&6.84$\pm$0.12&$-26.5\pm1.8$&$707\pm19$&$23.67\pm1.90$\\
\ion{He}{1}	&	6678	&--4.2$\pm$0.5	     &370$\pm$52 		&0.77$\pm$0.03&$-3.3\pm0.4$&$589\pm53$&$3.30\pm0.36$\\
\ion{H}{1}	&	7065	&--3.4$\pm$0.3	     &354$\pm$43		&0.67$\pm$0.04 &--&--&--\\
\ion{H}{1}	&	8446	&--2.1$\pm$0.5	     &313$\pm$36 		&0.42$\pm$0.04 &--&--&--\\
\ion{H}{1}	&	8500	&--1.5$\pm$0.1	     &362$\pm$19 		&0.34$\pm$0.03 &--&--&--\\
\ion{H}{1}	&	8543	&--2.2$\pm$0.3	     &401$\pm$51 		&0.48$\pm$0.06 &--&--&--\\
\ion{H}{1}	&	8597	&--2.0$\pm$0.4	     &377$\pm$40		&0.46$\pm$0.06 &--&--&--\\
\ion{H}{1}	&	8663	&--3.5$\pm$0.5	     &402$\pm$39		&0.76$\pm$0.05 &--&--&--\\
\ion{H}{1}	&	8749	&--3.7$\pm$0.5	     &437$\pm$30		&0.82$\pm$0.03 &--&--&--\\
\ion{H}{1}	&	8862	&--5.4$\pm$0.6	     &507$\pm$39		&1.17$\pm$0.14 &--&--&--\\
\ion{H}{1}	&	9015	&--3.8$\pm$0.7	     &420$\pm$51 		&0.83$\pm$0.34 &--&--&--\\
\ion{H}{1}	&	9228	&--14.1$\pm$1.0          &520$\pm$107 	&3.02$\pm$0.25 &--&--&--\\
\ion{H}{1}	&	9543	&--19.0$\pm$3.9          &515$\pm$36 		&4.23$\pm$0.29 &--&--&--\\
\ion{H}{1}	&	10050	&--14.6$\pm$3.2          &676$\pm$93 		&3.33$\pm$0.80 &--&--&--\\
\ion{He}{2}	&	10118	&--9.7$\pm$3.2	     &881$\pm$123 	&2.23$\pm$0.78 &--&--&--\\
\ion{H}{1}	&	10828	&--35.4$\pm$7.0          &606$\pm$100		&7.20$\pm$0.88 &--&--&--\\
\ion{H}{1}	&	10938	&--21.8$\pm$6.2          &652$\pm$113 	&4.39$\pm$0.68 &--&--&--\\
\hline
\end{tabular}
\end{minipage}
\end{table*}

\begin{table*}
\caption{Spectral Results for Bremsstrahlung Fits\label{tab:spec1}}
\begin{minipage}{\linewidth}
\begin{tabular}{ccccc} \hline \hline
Parameter\footnote{The errors on the parameters are 90\% confidence.} & Units & 1 Gaussian\footnote{The full model in XSPEC is {\ttfamily constant*tbabs*pcfabs*(gaussian+reflect*bremss)}.} & 2 Gaussians\footnote{This is the same model as the first column except for an additional Gaussian.  Ditto marks indicate parameters that are consistent with the first column.} & 3 Gaussians\footnote{This is the same model as the first column except for two additional Gaussians.  Ditto marks indicate parameters that are consistent with the other two columns.}\\ \hline
$N_{\rm H}$\footnote{The column density is calculated assuming \cite{wam00} abundances and \cite{vern96} cross sections.  Along this line of sight, the Galactic value is $N_{\rm H} = 1.8\times 10^{22}$\,cm$^{-2}$ \citep{kalberla05}.} & $10^{22}$\,cm$^{-2}$ & $0.45^{+0.09}_{-0.08}$ & '' & ''\\
$N_{\rm H,pc}$ & $10^{22}$\,cm$^{-2}$ & $8^{+3}_{-2}$ & '' & ''\\
pc fraction & --- & $0.65^{+0.04}_{-0.05}$ & '' & ''\\ \hline
$kT$ & keV & $81^{+31}_{-20}$ & '' & ''\\
$N_{\rm bremss}$\footnote{The normalization for the {\ttfamily bremss} model is equal to $\frac{3.02\times 10^{-15}}{4\pi d^{2}}\int{n_{e} n_{i} dV}$, where $d$ is the distance to the source in units of cm, $n_{e}$ and $n_{i}$ are the electron and ion densities in the plasma, and $V$ is the volume of the region containing the plasma.} & --- & $(4.4^{+0.4}_{-0.3})\times 10^{-4}$ & '' & ''\\ \hline
$\Omega/2\pi$ & --- & 1.0\footnote{Fixed.} & '' & ''\\
$A$\footnote{The abundance of elements heavier than He relative to solar.} & --- & $0.24^{+0.33}_{-0.16}$ & '' & ''\\
$A_{\rm Fe}$\footnote{The abundance of iron relative to the abundances specified by $A$.} & --- & 1.0$^{g}$ & '' & ''\\
$\cos{i}$ & --- & $>$0.70 & '' & ''\\ \hline
$E_{\rm line1}$ & keV & $6.59\pm 0.04$ & 6.4$^{g}$ & 6.4$^{g}$\\
$\sigma_{\rm line1}$ & keV & $0.28\pm 0.04$ & $0.24\pm 0.05$ & $0.07\pm 0.04$\\
$N_{\rm line1}$ & ph\,cm$^{-2}$\,s$^{-1}$ & $(2.3\pm 0.3)\times 10^{-5}$ & $(0.86^{+0.27}_{-0.29})\times 10^{-5}$ & $(1.0\pm 0.2)\times 10^{-5}$\\
$EW_{\rm line1}$ & eV & $940\pm 120$ & $230\pm 80$ & $320\pm 60$\\ \hline
$E_{\rm line2}$ & keV & --- & 6.7$^{g}$ & 6.7$^{g}$\\
$\sigma_{\rm line2}$ & keV & --- & 0.24\footnote{Tied to $\sigma_{\rm line1}$.} & 0.07$^{j}$\\
$N_{\rm line2}$ & ph\,cm$^{-2}$\,s$^{-1}$ & --- & $(1.4\pm 0.3)\times 10^{-5}$ & $(0.65\pm 0.16)\times 10^{-5}$\\
$EW_{\rm line2}$ & eV & --- &  $470\pm 100$ & $160\pm 40$\\ \hline
$E_{\rm line3}$ & keV & --- & --- & 6.97$^{g}$\\
$\sigma_{\rm line3}$ & keV & --- & --- & 0.07$^{j}$\\
$N_{\rm line3}$ & ph\,cm$^{-2}$\,s$^{-1}$ & --- & --- & $(0.51\pm 0.15)\times 10^{-5}$\\
$EW_{\rm line3}$ & eV & --- & --- & $170\pm 50$\\ \hline
$C_{\rm pn}$ & --- & 1.0$^{g}$ & '' & ''\\
$C_{\rm MOS1}$ & --- & $0.96\pm 0.05$ & '' & ''\\
$C_{\rm MOS2}$ & --- & $1.01\pm 0.05$ & '' & ''\\
$C_{\rm FPMA}$ & --- & $1.11\pm 0.07$ & '' & ''\\
$C_{\rm FPMB}$ & --- & $1.18\pm 0.07$ & '' & ''\\ \hline
$\chi^{2}$/dof & --- & 472/339 & 473/339 & 458/338\\ \hline
\end{tabular}
\end{minipage}
\end{table*}

\begin{table*}
\caption{Spectral Results for IP Mass Model Fits\label{tab:spec2}}
\begin{minipage}{\linewidth}
\begin{tabular}{ccc} \hline \hline
Parameter\footnote{The errors on the parameters are 90\% confidence.} & Units &  Value\footnote{The full model is {\ttfamily constant*tbabs*pcfabs*(gaussian+gaussian+gaussian+reflect*ipm)}.}\\ \hline
$N_{\rm H}$\footnote{The column density is calculated assuming \cite{wam00} abundances and \cite{vern96} cross sections.  Along this line of sight, the Galactic value is $N_{\rm H} = 1.8\times 10^{22}$\,cm$^{-2}$ \citep{kalberla05}.} & $10^{22}$\,cm$^{-2}$ & $0.63^{+0.09}_{-0.08}$\\
$N_{\rm H,pc}$ & $10^{22}$\,cm$^{-2}$ & $12^{+3}_{-2}$\\
pc fraction & --- & $0.72\pm 0.03$\\ \hline
$M_{\rm WD}$ & \Msun & $>$1.38\\
$N_{\rm IPM}$ & --- & $(1.63^{+0.23}_{-0.68})\times 10^{-13}$\\ \hline
$\Omega/2\pi$ & --- & 1.0\footnote{Fixed.}\\
$A$\footnote{The abundance of elements heavier than He relative to solar.} & --- & $0.4^{+0.5}_{-0.2}$\\
$A_{\rm Fe}$\footnote{The abundance of iron relative to the abundances specified by $A$.} & --- & 1.0$^{d}$\\
$\cos{i}$ & --- & $>$0.84\\ \hline
$E_{\rm line1}$ & keV & 6.4$^{d}$\\
$\sigma_{\rm line1}$ & keV & $0.07^{+0.04}_{-0.03}$\\
$N_{\rm line1}$ & ph\,cm$^{-2}$\,s$^{-1}$ & $(1.1\pm 0.2)\times 10^{-5}$\\ \hline
$E_{\rm line2}$ & keV & 6.7$^{d}$\\
$\sigma_{\rm line2}$ & keV & 0.07\footnote{Tied to $\sigma_{\rm line1}$.}\\
$N_{\rm line2}$ & ph\,cm$^{-2}$\,s$^{-1}$ & $(0.67\pm 0.16)\times 10^{-5}$\\ \hline
$E_{\rm line3}$ & keV & 6.97$^{d}$\\
$\sigma_{\rm line3}$ & keV & 0.07$^{g}$\\
$N_{\rm line3}$ & ph\,cm$^{-2}$\,s$^{-1}$ & $(0.51\pm 0.15)\times 10^{-5}$\\ \hline
$C_{\rm pn}$ & --- & 1.0$^{d}$\\
$C_{\rm MOS1}$ & --- & $0.96\pm 0.05$\\
$C_{\rm MOS2}$ & --- & $1.00\pm 0.05$\\
$C_{\rm FPMA}$ & --- & $1.13\pm 0.07$\\
$C_{\rm FPMB}$ & --- & $1.20\pm 0.07$\\ \hline
$\chi^{2}$/dof & --- & 480/338\\ \hline
\end{tabular}
\end{minipage}
\end{table*}

\section*{Acknowledgments}

We would like to thank D. Wik for help with using {\ttfamily nuskybgd} to produce a
background spectrum for {\em NuSTAR}.  We also thank C. Hailey, J. Hong, and 
F. Fornasini for useful discussions.  FR thanks the ESO staff who performed the 
service observations.  
JAT acknowledges partial support from NASA under {\em XMM} Guest Observer grant 
NNX15AW09G.  MC acknowledges partial support under NASA Contract NNG08FD60C 
for work on the {\em NuSTAR} mission.  This work was partially supported by 
NASA {\em Fermi} grant NNX15AU83G.  RK acknowledges support from Russian Science 
Foundation (grant 14-22-00271).  
This work made use of data from the {\it NuSTAR} mission, a project led by the 
California Institute of Technology, managed by the Jet Propulsion Laboratory, 
and funded by the National Aeronautics and Space Administration. We thank the 
{\it NuSTAR} Operations, Software and  Calibration teams for support with the 
execution and analysis of these observations.  This research has made use of 
the {\it NuSTAR} Data Analysis Software (NuSTARDAS) jointly developed by the 
ASI Science Data Center (ASDC, Italy) and the California Institute of Technology 
(USA).  Based on observations obtained at the Southern Astrophysical Research 
(SOAR) telescope, which is a joint project of the Minist\'{e}rio da Ci\^{e}ncia, 
Tecnologia, e Inova\c{c}\~{a}o (MCTI) da Rep\'{u}blica Federativa do Brasil, the 
U.S. National Optical Astronomy Observatory (NOAO), the University of North 
Carolina at Chapel Hill (UNC), and Michigan State University (MSU).\\



\begin{thebibliography}{48}
\expandafter\ifx\csname natexlab\endcsname\relax\def\natexlab#1{#1}\fi

\bibitem[{{Arnaud}(1996)}]{arnaud96}
{Arnaud} K.~A., 1996, in Astronomical Society of the Pacific Conference Series,
  Vol. 101, Astronomical Data Analysis Software and Systems V, {Jacoby} G.~H.,
  {Barnes} J., eds., p.~17

\bibitem[{{Beardmore} {et~al}\mbox{.}(2000){Beardmore}, {Osborne}, \&
  {Hellier}}]{boh00}
{Beardmore} A.~P., {Osborne} J.~P., {Hellier} C., 2000, MNRAS, 315, 307

\bibitem[{{Belle} {et~al}\mbox{.}(2003){Belle}, {Howell}, {Sion}, {Long}, \&
  {Szkody}}]{belle03}
{Belle} K.~E., {Howell} S.~B., {Sion} E.~M., {Long} K.~S., {Szkody} P., 2003,
  ApJ, 587, 373

\bibitem[{{Bird} {et~al}\mbox{.}(2016){Bird}, {Bazzano}, {Malizia}, {Fiocchi},
  {Sguera}, {Bassani}, {Hill}, {Ubertini}, \& {Winkler}}]{bird16}
{Bird} A.~J. {et~al.}, 2016, arXiv:1601.06074

\bibitem[{{Buccheri} {et~al}\mbox{.}(1983){Buccheri}, {Bennett}, {Bignami},
  {Bloemen}, {Boriakoff}, {Caraveo}, {Hermsen}, {Kanbach}, {Manchester},
  {Masnou}, {Mayer-Hasselwander}, {Ozel}, {Paul}, {Sacco}, {Scarsi}, \&
  {Strong}}]{buccheri83}
{Buccheri} R. {et~al.}, 1983, A\&A, 128, 245

\bibitem[{{Clemens} {et~al}\mbox{.}(2004){Clemens}, {Crain}, \&
  {Anderson}}]{clemens04}
{Clemens} J.~C., {Crain} J.~A., {Anderson} R., 2004, in Society of
  Photo-Optical Instrumentation Engineers (SPIE) Conference Series, Vol. 5492,
  Ground-based Instrumentation for Astronomy, {Moorwood} A.~F.~M., {Iye} M.,
  eds., pp. 331--340

\bibitem[{{Di Stefano}(2010{\natexlab{a}})}]{distefano10a}
{Di Stefano} R., 2010{\natexlab{a}}, ApJ, 712, 728

\bibitem[{{Di Stefano}(2010{\natexlab{b}})}]{distefano10b}
{Di Stefano} R., 2010{\natexlab{b}}, ApJ, 719, 474

\bibitem[{{Gilfanov} \& {Bogd{\'a}n}(2010)}]{gb10}
{Gilfanov} M., {Bogd{\'a}n} {\'A}., 2010, Nature, 463, 924

\bibitem[{{Graur} {et~al}\mbox{.}(2014){Graur}, {Maoz}, \& {Shara}}]{graur14}
{Graur} O., {Maoz} D., {Shara} M.~M., 2014, MNRAS, 442, L28

\bibitem[{{G{\"u}ver} \& {{\"O}zel}(2009)}]{go09}
{G{\"u}ver} T., {{\"O}zel} F., 2009, MNRAS, 400, 2050

\bibitem[{{Harrison} {et~al}\mbox{.}(2013){Harrison}, {Craig}, {Christensen},
  {Hailey}, {Zhang}, {Boggs}, {Stern}, {Cook}, {Forster}, {Giommi},
  {Grefenstette}, {Kim}, {Kitaguchi}, {Koglin}, {Madsen}, {Mao}, {Miyasaka},
  {Mori}, {Perri}, {Pivovaroff}, {Puccetti}, {Rana}, {Westergaard}, {Willis},
  {Zoglauer}, {An}, {Bachetti}, {Barri{\`e}re}, {Bellm}, {Bhalerao},
  {Brejnholt}, {Fuerst}, {Liebe}, {Markwardt}, {Nynka}, {Vogel}, {Walton},
  {Wik}, {Alexander}, {Cominsky}, {Hornschemeier}, {Hornstrup}, {Kaspi},
  {Madejski}, {Matt}, {Molendi}, {Smith}, {Tomsick}, {Ajello}, {Ballantyne},
  {Balokovi{\'c}}, {Barret}, {Bauer}, {Blandford}, {Brandt}, {Brenneman},
  {Chiang}, {Chakrabarty}, {Chenevez}, {Comastri}, {Dufour}, {Elvis}, {Fabian},
  {Farrah}, {Fryer}, {Gotthelf}, {Grindlay}, {Helfand}, {Krivonos}, {Meier},
  {Miller}, {Natalucci}, {Ogle}, {Ofek}, {Ptak}, {Reynolds}, {Rigby},
  {Tagliaferri}, {Thorsett}, {Treister}, \& {Urry}}]{harrison13}
{Harrison} F.~A. {et~al.}, 2013, ApJ, 770, 103

\bibitem[{{Hayashi} {et~al}\mbox{.}(2011){Hayashi}, {Ishida}, {Terada},
  {Bamba}, \& {Shionome}}]{hayashi11}
{Hayashi} T., {Ishida} M., {Terada} Y., {Bamba} A., {Shionome} T., 2011, PASJ,
  63, S739

\bibitem[{{Hellier} \& {Mukai}(2004)}]{hm04}
{Hellier} C., {Mukai} K., 2004, MNRAS, 352, 1037

\bibitem[{{Hong} {et~al}\mbox{.}(2012){Hong}, {van den Berg}, {Grindlay},
  {Servillat}, \& {Zhao}}]{hong12}
{Hong} J., {van den Berg} M., {Grindlay} J.~E., {Servillat} M., {Zhao} P.,
  2012, ApJ, 746, 165

\bibitem[{{Jenniskens} \& {Desert}(1994)}]{1994Jenniskens}
{Jenniskens} P., {Desert} F.-X., 1994, A\&AS, 106, 39

\bibitem[{{Kalberla} {et~al}\mbox{.}(2005){Kalberla}, {Burton}, {Hartmann},
  {Arnal}, {Bajaja}, {Morras}, \& {P{\"o}ppel}}]{kalberla05}
{Kalberla} P.~M.~W., {Burton} W.~B., {Hartmann} D., {Arnal} E.~M., {Bajaja} E.,
  {Morras} R., {P{\"o}ppel} W.~G.~L., 2005, A\&A, 440, 775

\bibitem[{{Krivonos} {et~al}\mbox{.}(2007){Krivonos}, {Revnivtsev}, {Churazov},
  {Sazonov}, {Grebenev}, \& {Sunyaev}}]{krivonos07_cv}
{Krivonos} R., {Revnivtsev} M., {Churazov} E., {Sazonov} S., {Grebenev} S.,
  {Sunyaev} R., 2007, A\&A, 463, 957

\bibitem[{{Krivonos} {et~al}\mbox{.}(2012){Krivonos}, {Tsygankov}, {Lutovinov},
  {Revnivtsev}, {Churazov}, \& {Sunyaev}}]{krivonos12}
{Krivonos} R., {Tsygankov} S., {Lutovinov} A., {Revnivtsev} M., {Churazov} E.,
  {Sunyaev} R., 2012, A\&A, 545, A27

\bibitem[{{Landi} {et~al}\mbox{.}(2012){Landi}, {Bassani}, {Masetti},
  {Bazzano}, {Fiocchi}, {Bird}, \& {Drave}}]{landi12}
{Landi} R., {Bassani} L., {Masetti} N., {Bazzano} A., {Fiocchi} M., {Bird}
  A.~J., {Drave} S., 2012, The Astronomer's Telegram, 4165, 1

\bibitem[{{Liu} {et~al}\mbox{.}(2012){Liu}, {Di Stefano}, {Wang}, \&
  {Moe}}]{liu12}
{Liu} J., {Di Stefano} R., {Wang} T., {Moe} M., 2012, ApJ, 749, 141

\bibitem[{{Magdziarz} \& {Zdziarski}(1995)}]{mz95}
{Magdziarz} P., {Zdziarski} A.~A., 1995, MNRAS, 273, 837

\bibitem[{{Marshall} {et~al}\mbox{.}(2006){Marshall}, {Robin}, {Reyl{\'e}},
  {Schultheis}, \& {Picaud}}]{marshall06}
{Marshall} D.~J., {Robin} A.~C., {Reyl{\'e}} C., {Schultheis} M., {Picaud} S.,
  2006, A\&A, 453, 635

\bibitem[{{Mukai} {et~al}\mbox{.}(2015){Mukai}, {Rana}, {Bernardini}, \& {de
  Martino}}]{mukai15}
{Mukai} K., {Rana} V., {Bernardini} F., {de Martino} D., 2015, ApJ, 807, L30

\bibitem[{{Nielsen} {et~al}\mbox{.}(2013{\natexlab{a}}){Nielsen}, {Dominik},
  {Nelemans}, \& {Voss}}]{nielsen13a}
{Nielsen} M.~T.~B., {Dominik} C., {Nelemans} G., {Voss} R., 2013{\natexlab{a}},
  A\&A, 549, A32

\bibitem[{{Nielsen} {et~al}\mbox{.}(2014){Nielsen}, {Gilfanov}, {Bogd{\'a}n},
  {Woods}, \& {Nelemans}}]{nielsen14}
{Nielsen} M.~T.~B., {Gilfanov} M., {Bogd{\'a}n} {\'A}., {Woods} T.~E.,
  {Nelemans} G., 2014, MNRAS, 442, 3400

\bibitem[{{Nielsen} {et~al}\mbox{.}(2012){Nielsen}, {Voss}, \&
  {Nelemans}}]{nielsen12}
{Nielsen} M.~T.~B., {Voss} R., {Nelemans} G., 2012, MNRAS, 426, 2668

\bibitem[{{Nielsen} {et~al}\mbox{.}(2013{\natexlab{b}}){Nielsen}, {Voss}, \&
  {Nelemans}}]{nielsen13b}
{Nielsen} M.~T.~B., {Voss} R., {Nelemans} G., 2013{\natexlab{b}}, MNRAS, 435,
  187

\bibitem[{{Perez} {et~al}\mbox{.}(2015){Perez}, {Hailey}, {Bauer}, {Krivonos},
  {Mori}, {Baganoff}, {Barri{\`e}re}, {Boggs}, {Christensen}, {Craig},
  {Grefenstette}, {Grindlay}, {Harrison}, {Hong}, {Madsen}, {Nynka}, {Stern},
  {Tomsick}, {Wik}, {Zhang}, {Zhang}, \& {Zoglauer}}]{perez15}
{Perez} K. {et~al.}, 2015, Nature, 520, 646

\bibitem[{{Rahoui} {et~al}\mbox{.}(2014){Rahoui}, {Coriat}, \&
  {Lee}}]{2014Rahoui}
{Rahoui} F., {Coriat} M., {Lee} J.~C., 2014, MNRAS, 442, 1610

\bibitem[{{Ritter} \& {Kolb}(2003)}]{rk03}
{Ritter} H., {Kolb} U., 2003, A\&A, 404, 301

\bibitem[{{Scaringi} {et~al}\mbox{.}(2011){Scaringi}, {Connolly}, {Patterson},
  {Thorstensen}, {Uthas}, {Knigge}, {Vican}, {Monard}, {Rea}, {Krajci},
  {Lowther}, {Myers}, {Bolt}, {Dieball}, \& {Groot}}]{scaringi11}
{Scaringi} S. {et~al.}, 2011, A\&A, 530, A6

\bibitem[{{Still} {et~al}\mbox{.}(1998){Still}, {Duck}, \& {Marsh}}]{still98}
{Still} M.~D., {Duck} S.~R., {Marsh} T.~R., 1998, MNRAS, 299, 759

\bibitem[{{Str{\"u}der} {et~al}\mbox{.}(2001){Str{\"u}der}, {Briel}, {Dennerl},
  {Hartmann}, {Kendziorra}, {Meidinger}, {Pfeffermann}, {Reppin}, {Aschenbach},
  {Bornemann}, {Br{\"a}uninger}, {Burkert}, {Elender}, {Freyberg}, {Haberl},
  {Hartner}, {Heuschmann}, {Hippmann}, {Kastelic}, {Kemmer}, {Kettenring},
  {Kink}, {Krause}, {M{\"u}ller}, {Oppitz}, {Pietsch}, {Popp}, {Predehl},
  {Read}, {Stephan}, {St{\"o}tter}, {Tr{\"u}mper}, {Holl}, {Kemmer}, {Soltau},
  {St{\"o}tter}, {Weber}, {Weichert}, {von Zanthier}, {Carathanassis}, {Lutz},
  {Richter}, {Solc}, {B{\"o}ttcher}, {Kuster}, {Staubert}, {Abbey}, {Holland},
  {Turner}, {Balasini}, {Bignami}, {La Palombara}, {Villa}, {Buttler},
  {Gianini}, {Lain{\'e}}, {Lumb}, \& {Dhez}}]{struder01}
{Str{\"u}der} L. {et~al.}, 2001, A\&A, 365, L18

\bibitem[{{Suleimanov} {et~al}\mbox{.}(2005){Suleimanov}, {Revnivtsev}, \&
  {Ritter}}]{srr05}
{Suleimanov} V., {Revnivtsev} M., {Ritter} H., 2005, A\&A, 435, 191

\bibitem[{{Taylor} {et~al}\mbox{.}(1997){Taylor}, {Beardmore}, {Norton},
  {Osborne}, \& {Watson}}]{taylor97}
{Taylor} P., {Beardmore} A.~P., {Norton} A.~J., {Osborne} J.~P., {Watson}
  M.~G., 1997, MNRAS, 289, 349

\bibitem[{{Tomsick} {et~al}\mbox{.}(2015){Tomsick}, {Krivonos}, {Rahoui},
  {Ajello}, {Rodriguez}, {Barri{\`e}re}, {Bodaghee}, \& {Chaty}}]{tomsick15}
{Tomsick} J.~A., {Krivonos} R., {Rahoui} F., {Ajello} M., {Rodriguez} J.,
  {Barri{\`e}re} N., {Bodaghee} A., {Chaty} S., 2015, MNRAS, 449, 597

\bibitem[{{Tomsick} {et~al}\mbox{.}(2016){Tomsick}, {Krivonos}, {Wang},
  {Bodaghee}, {Chaty}, {Rahoui}, {Rodriguez}, \& {Fornasini}}]{tomsick16}
{Tomsick} J.~A., {Krivonos} R., {Wang} Q., {Bodaghee} A., {Chaty} S., {Rahoui}
  F., {Rodriguez} J., {Fornasini} F.~M., 2016, ApJ, 816, 38

\bibitem[{{Tomsick} {et~al}\mbox{.}(2014){Tomsick}, {Rahoui}, {Krivonos},
  {Rodriguez}, {Bodaghee}, \& {Chaty}}]{tomsick14_atel}
{Tomsick} J.~A., {Rahoui} F., {Krivonos} R., {Rodriguez} J., {Bodaghee} A.,
  {Chaty} S., 2014, The Astronomer's Telegram, 6793, 1

\bibitem[{{Turner} {et~al}\mbox{.}(2001){Turner}, {Abbey}, {Arnaud},
  {Balasini}, {Barbera}, {Belsole}, {Bennie}, {Bernard}, {Bignami}, {Boer},
  {Briel}, {Butler}, {Cara}, {Chabaud}, {Cole}, {Collura}, {Conte}, {Cros},
  {Denby}, {Dhez}, {Di Coco}, {Dowson}, {Ferrando}, {Ghizzardi}, {Gianotti},
  {Goodall}, {Gretton}, {Griffiths}, {Hainaut}, {Hochedez}, {Holland},
  {Jourdain}, {Kendziorra}, {Lagostina}, {Laine}, {La Palombara}, {Lortholary},
  {Lumb}, {Marty}, {Molendi}, {Pigot}, {Poindron}, {Pounds}, {Reeves},
  {Reppin}, {Rothenflug}, {Salvetat}, {Sauvageot}, {Schmitt}, {Sembay},
  {Short}, {Spragg}, {Stephen}, {Str{\"u}der}, {Tiengo}, {Trifoglio},
  {Tr{\"u}mper}, {Vercellone}, {Vigroux}, {Villa}, {Ward}, {Whitehead}, \&
  {Zonca}}]{turner01}
{Turner} M.~J.~L. {et~al.}, 2001, A\&A, 365, L27

\bibitem[{{Verner} {et~al}\mbox{.}(1996){Verner}, {Ferland}, {Korista}, \&
  {Yakovlev}}]{vern96}
{Verner} D.~A., {Ferland} G.~J., {Korista} K.~T., {Yakovlev} D.~G., 1996, ApJ,
  465, 487

\bibitem[{{Wik} {et~al}\mbox{.}(2014){Wik}, {Hornstrup}, {Molendi}, {Madejski},
  {Harrison}, {Zoglauer}, {Grefenstette}, {Gastaldello}, {Madsen},
  {Westergaard}, {Ferreira}, {Kitaguchi}, {Pedersen}, {Boggs}, {Christensen},
  {Craig}, {Hailey}, {Stern}, \& {Zhang}}]{wik14}
{Wik} D.~R. {et~al.}, 2014, ApJ, 792, 48

\bibitem[{{Williams} \& {Shipman}(1988)}]{1988Williams}
{Williams} G.~A., {Shipman} H.~L., 1988, ApJ, 326, 738

\bibitem[{{Williams}(1980)}]{1980Williams}
{Williams} R.~E., 1980, ApJ, 235, 939

\bibitem[{{Wilms} {et~al}\mbox{.}(2000){Wilms}, {Allen}, \& {McCray}}]{wam00}
{Wilms} J., {Allen} A., {McCray} R., 2000, ApJ, 542, 914

\bibitem[{{Winkler} {et~al}\mbox{.}(2003){Winkler}, {Courvoisier}, {Di Cocco},
  {Gehrels}, {Gim{\' e}nez}, {Grebenev}, {Hermsen}, {Mas-Hesse}, {Lebrun},
  {Lund}, {Palumbo}, {Paul}, {Roques}, {Schnopper}, {Sch{\" o}nfelder},
  {Sunyaev}, {Teegarden}, {Ubertini}, {Vedrenne}, \& {Dean}}]{winkler03}
{Winkler} C. {et~al.}, 2003, A\&A, 411, L1

\bibitem[{{Yuasa} {et~al}\mbox{.}(2010){Yuasa}, {Nakazawa}, {Makishima},
  {Saitou}, {Ishida}, {Ebisawa}, {Mori}, \& {Yamada}}]{yuasa10}
{Yuasa} T., {Nakazawa} K., {Makishima} K., {Saitou} K., {Ishida} M., {Ebisawa}
  K., {Mori} H., {Yamada} S., 2010, A\&A, 520, A25

\bibitem[{{Zorotovic} {et~al}\mbox{.}(2011){Zorotovic}, {Schreiber}, \&
  {G{\"a}nsicke}}]{zsg11}
{Zorotovic} M., {Schreiber} M.~R., {G{\"a}nsicke} B.~T., 2011, A\&A, 536, A42

\end{thebibliography}

\label{lastpage}
\bsp

\end{document}